\documentclass[a4paper,11pt]{article}

\usepackage{graphicx}
\usepackage{dcolumn}
\usepackage{bm}
\pdfoutput=1
\usepackage{amssymb}
\usepackage{amsfonts}
\usepackage{amsmath}

\newcommand\nc\newcommand
\nc\bit{\begin{itemize}}
\nc\eit{\end{itemize}}
\nc\noi{\noindent}
\nc\be{\begin{equation}}
\nc\ee{\end{equation}}
\nc \ba{\begin{eqnarray}}
\nc\ea{\end{eqnarray}}
\nc \bastar{\begin{eqnarray*}}
\nc\eastar{\end{eqnarray*}}
\nc \barr{\begin{array}}
\nc\earr{\end{array}}

\def\qed{\ifmmode{\hbox{\rlap{$\sqcap$}$\sqcup$}}\else{\unskip\nobreak\hfil
\penalty50\hskip1em\null\nobreak\hfil{\hbox{\rlap{$\sqcap$}$\sqcup$}}
\parfillskip=0pt\finalhyphendemerits=0\endgraf}\fi}

\nc\Preuve{\removelastskip\vskip\baselineskip\noindent{\it Preuve.\quad}\ignorespaces}
 \nc\Preuvede[1]{\removelastskip\vskip\baselineskip\noindent{\it Preuve de la {#1}. \quad}\ignorespaces}
 \nc\Preuvedu[1]{\removelastskip\vskip\baselineskip\noindent{\it Preuve du {#1}. \quad}\ignorespaces}
\nc\Proof{\removelastskip\vskip\baselineskip\noindent{\it Proof.\quad}\ignorespaces}
 \nc\Proofof[1]{\removelastskip\vskip\baselineskip\noindent{\it Proof of {#1}.
\quad}\ignorespaces}

\newcounter{enonce}[section]

\newcounter{theor}[section]

\newcounter{propos}[section]

\newcounter{example}[section]

\newcounter{remark}[section]

\newcounter{defini}[section]

\newcounter{lemm}[section]

\newcounter{corol}[section]

\newcounter{condi}[section]

\newcounter{methode}[section]

\newcounter{exemple}[section]

\newcounter{theora}[section]

\newcounter{definia}[section]

\newcounter{lemma}[section]

\newcounter{corola}[section]

\newcounter{hypo}[section]

\nc\chap[1]{\setcounter{equation}{0}\chapter{#1}}
\nc\sect[1]{\setcounter{equation}{0} \section{#1}}
\nc\subsect[1]{\setcounter{equation}{0}\subsection{#1}}

\nc\conj[1]{\overline{#1}}
\nc\der[2]{{\partial{#1}\over\partial{#2}}}
\nc\ra{{\rightarrow}}
\nc\Ra{{\Rightarrow}}
\nc\lra{{\longrightarrow}}
\nc\Lra{{\Longrightarrow}}
\nc\Lera{\Leftrightarrow}
\nc\Llra{\Longleftrightarrow}

\def\abs#1{\left\vert #1 \right\vert} 
\def\norme#1{\left\Vert #1 \right\Vert} 
\nc\limm{\mathrm{l.i.m}\ }

\def\cali#1{\cal#1\mit}
\nc\slim{\mathrm{s.}\negthinspace\lim}
\nc\LL{ L^2(\mathbb{R})}
\nc\ran{{\rm Ran}}
\nc\limlog[2]{\frac{\log\left(#1\right)}{\log(#2)}}
\nc\dda[1]{\frac{d #1}{#1}}
\nc\inv[1]{\frac{1}{#1}}

\nc\PP{\mathrm{Proba}}
\nc\RR{\mathbb{R}}
\nc\RRstar{\mathbb{R}\backslash 0}
\nc\CC{\mathbb{C}}
\nc\QQ{\mathbb{Q}}
\nc\NN{\mathbb{N}}
\nc\ZZ{\mathbb{Z}}
\nc\II{\mathbb{I}}
\nc\HH{\mathbb{H}}
\nc\EE{\mathbb{E}}
\nc\diag{{\mathbf{Diag}}}
\nc\SSS{\cali S(\RR)}
\nc\SSSp{\cali S'(\RR)}
\nc\vect[1]{\boldsymbol{#1}}
\nc\mean[1]{\overline{\boldsymbol{#1}}}

\nc\tens[1]{\boldsymbol{\mathbb{{#1}}}}
\nc\arpp[1]{\tens A\tens C^+(\vect{#1})}
\nc\arpm[1]{\tens T(\vect{#1})}

\nc\dir[1]{\boldsymbol{\widehat #1}}
\nc\scal[2]{\boldsymbol{#1}\cdot \boldsymbol{#2}}

\nc\diam[1]{\mbox{diam(#1)}}
\nc\var[1]{\mbox{Var}(#1)}
\nc\Cor[1]{\mbox{Cor}(#1)}
\nc\curl{\mathrm{curl}}
\nc\refer[1]{{#1}_{\mathrm{ref}}}
\nc\vectrefer[1]{{\boldsymbol{#1}}_{\mathrm{ref}}}

\nc\const{\mathrm{const}\times}
\nc\normeop[1]{\abs{\norme{#1}}}

\newcommand{\beq}[1]{\begin{equation}\label{#1}}
\newcommand{\eeq}{\end{equation}}
\newcommand{\bay}[1]{\begin{eqnarray}\label{#1}}
\newcommand{\eay}{\end{eqnarray}}
\newcommand{\bsa}[1]{\begin{subeqnarray}\label{#1}}
\newcommand{\esa}{\end{subeqnarray}}
\newcommand{\bparts}[1]{\begin{eqnarray}\label{#1}\numparts}
\newcommand{\eparts}{\endnumparts \end{eqnarray}}

\newsavebox{\astrutbox}
\sbox{\astrutbox}{\rule[-5pt]{0pt}{20pt}}

\begin{document}

\title{Self-calibration and antenna grouping for bistatic oceanographic High-Frequency Radars}
\author{Dylan DUMAS and Charles-Antoine GU\'ERIN\\
{\small University of Toulon and Mediterranean Institute of Oceanography}\\
{\small (MIO UM 110, Univ Toulon, Aix-Marseille Univ, CNRS, IRD)}\\
{\small La Garde, France}\\
{\small dylan.dumas@univ-tln.fr, guerin@univ-tln.fr}}

\date{\today}

\maketitle
\begin{abstract}
  We propose two concepts for the significant improvement of surface current mapping with bistatic oceanographic High-Frequency Radars. These ameliorations pertain to the azimuthal processing of radar data with linear or quasi-linear antenna arrays. The first idea is to take advantage of the remote transmitter to perform an automatic correction of the complex gains of the receiving antennas based on the analysis of the signal received in the direct path. This direct signal can be found at the zero-Doppler and minimal range cell in the Range-Doppler representation. We term this adjustment as ``self-calibration'' of the receiving array, as it can be performed in real-time without any specific action from the operator. The second idea consists in applying a Direction Finding technique (instead of traditional Beam Forming) not only to the full array of antenna but also to subarrays made of a smaller number of sequential antennas, a method which we refer to as ``antenna grouping''. The combination of self-calibration and antenna grouping makes it possible to obtain high-resolution maps with full coverage, thereby combining the respective merits of Direction Finding and Beam Forming techniques. In addition the method is found robust to missing antennas in the array. These techniques are applied to and illustrated with the multistatic High-Frequency Radar network in Toulon.
\end{abstract}
\section{Introduction}

High-Frequency Radars (HFR) are routinely used for the mapping of coastal surface currents. The main physical principle underlying this detection have been unveiled in the pioneering works of Crombie \cite{crombie_Nature55} and Barrick \cite{barrick_72conf1}. It relies on measuring the Doppler shift induced by the radial surface current on the backscattered sea echo. This is made possible by the presence of a couple of very marked peaks in the Doppler spectrum, referred to as the ``Bragg lines''. This terminology alludes to a grating effect which is observed in HFR scattering from gravity waves and which is similar to the resonant mechanism observed in X-ray Bragg diffraction from crystals. There is abundant literature on the estimation of surface currents from HFR and the associated applications (see e.g. the review papers \cite{lipa1981,paduan1997,headrick1998,paduan2013,wyatt2014,roarty2019}) and we will not enter in the details. Suffice to say, the extraction of surface current maps from HFR antenna voltage results from a complex but universal chain of processing whose main steps can be synthesized as follows:
\begin{enumerate}
\item For each receiving antenna, record a coherent complex time series (I,Q channels) resulting from the echo of a succession of emitted chirps from the transmitting antenna.
\item For each sweep and antenna, process the time series in range by performing a Fast Fourier Transform (range processing).
\item For each antenna and range-resolved time series at the sweep rate, perform another Fast Fourier Transform to obtain the complex Doppler spectrum (Doppler processing).
\item Combine the range-resolved Doppler spectra from all antennas to determine the bearings of surface current projections (azimuthal processing).
\item Combine the range-azimuth resolved surface current projections to obtain the surface current vector map (total current reconstruction).
\end{enumerate}

\begin{figure}[htbp]\centering
	  \includegraphics[scale=0.5]{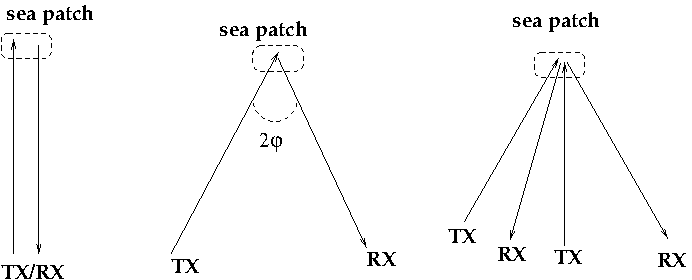}
\caption{Monostatic (left), bistatic (middle) and multistatic (right) configuration. The bistatic angle $\varphi$ is the key parameter of the bistatic configuration.}
\label{fig1}
\end{figure}

Most HFR systems are monostatic, that is have co-located transmitters and receivers, and use pairs of radar to infer two radial components from which the surface current vector can be recombined. In some cases it is advantageous to operate in bistatic configuration (e.g. \cite{grosdidier_GRS14})  in which the transmitter and receiver are located remotely (see Figure \ref{fig1}) or even multistatic systems which are a combination of remote  transmitters and receivers (e.g. \cite{dumas_OD20}).

The aforementioned chain of processing is common to monostatic and bistatic configuration with some adaptation in the geometrical formulas. For a bistatic pair of (TX,RX) the iso-range radar cells follow ellipses with focii at the transmitter and receiver locations (as opposed to circles around the transmitter in the monostatic case). The resonant Bragg frequency $f_B$ in the Doppler spectrum depends on the bistatic angle $\varphi$ (see Figure \ref{fig1}), hence on the sea surface patch ($f_B^2=g\cos\varphi/(\pi\lambda)$) while it is constant in the monostatic case ($f_B^2=g/(\pi\lambda)$). Any observed Doppler shift $\Delta f$ with respect to this local Bragg frequency is proportional to the projected component $U_n$ of the surface current vector $\vect U$ onto the normal direction to the ellipse, which is referred to as the elliptical velocity ($U_n=\lambda \Delta f/(2\cos\varphi)$). In the monostatic case this frequency shift is proportional to the radial component $U_r$ of the surface current vector along the radar look direction,  ($U_r=\lambda\Delta f/2$).

The most critical and system-dependent operation in processing HF radar data for producing current maps is the azimuthal discrimination of the received signal (step 4 in the chain of processing). For linear or quasi-linear extended arrays of antennas, this is usually done with a Beam Forming (BF) technique which allows to steer the bearing angle by adjusting numerically the relative phase shifts of the antenna signals. This makes it possible to sweep continuously the angular sector covered by the radar. However, the resulting azimuthal accuracy depends on the array extension and deteriorates significantly as the steering angle deviates from the central direction. Compact antenna systems rely on high-resolution methods such as Direction Finding (DF) techniques with the weak point that this requires longer integration time and produces lacunary maps. These two techniques will be analyzed and compared in the following in the context of the HFR network in Toulon. This last system has been operated for one decade in bistatic mode and is running in multistatic mode as of January 2019 (\cite{grosdidier_GRS14,guerin_radar2019,dumas_OD20}). While developing specific software for the azimuthal processing of these data in such a nonstandard configuration we discovered a novel opportunity offered by the bistatic mode: the possibility to calibrate the receiving antenna with the direct signal of the remote transmitter without resorting to any other traditional technique of calibration (typically, ship calibration). We refer to this simple and automatic technique as a ``self-calibration'' method since it does not require any specific action from the operator (besides an additional line of code in the software) and can be performed in real time for every time series. Another original aspect of the historical network was the limited extension and irregular form of the initial antenna arrays (8 antenna from 2012 to 2018) which conducted its first operators to develop high-resolution DF methods (\cite{barbin09,barbin11}). The antenna arrays were extended to linear arrays of 12 antennas in 2019 and 2020 but the idea of using DF for non-compact arrays was maintained and improved. In 2019 we devised an improved Direction-Finding technique based on testing multiple subarrays, a method which we termed ``antenna grouping''. The self-calibration and antenna grouping techniques are currently running on site for the processing of near real-time HFR data in Toulon (http://hfradar.univ-tln.fr/) and have been used for the reprocessing of the historical data (2012-2018). After a brief presentation of this HFR network (Section \ref{sec:HFR}), we will introduce the self-calibration method in the context of BF (Sections \ref{sec:BF} and \ref{sec:sc}) and the antenna grouping in the context of DF (Sections \ref{sec:DF} and \ref{sec:ag}). When used together, these methods allow for a significant improvement of surface current mapping in terms of accuracy, coverage and robustness to hardware failure.

\section{The HFR in Toulon}\label{sec:HFR}

\begin{figure}[htbp]\centering
	  \includegraphics[scale=0.3]{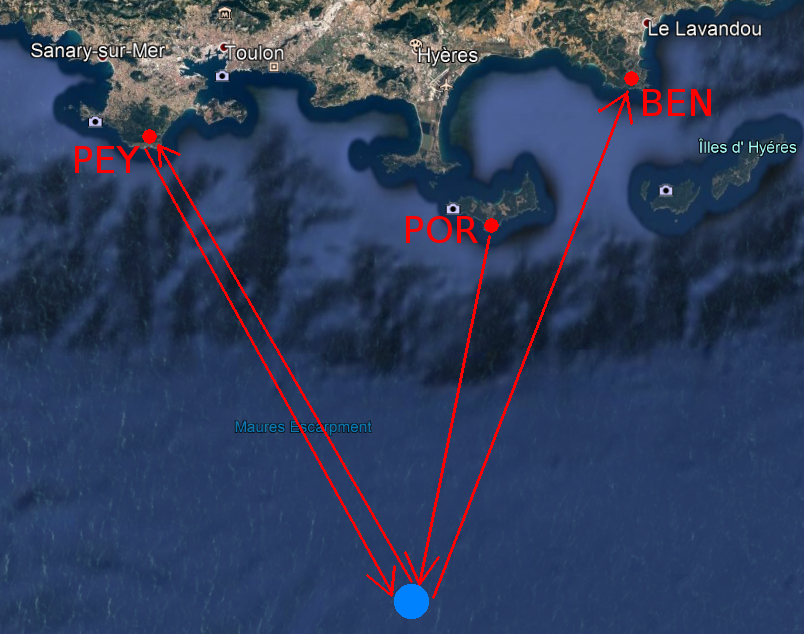}
\caption{The three HFR sites in the region of Toulon: 1) Fort Peyras (TX/RX, ``PEY''); 2) Cap B\'enat (RX, ``BEN''); 3) Porquerolles Island (TX, ``POR'').}
\label{fig2}
\end{figure}

\begin{figure}[htbp]
  \centering
	 \includegraphics[scale=0.75]{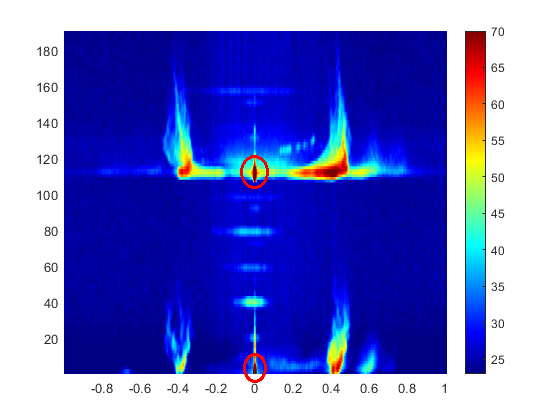}
	\includegraphics[scale=0.75]{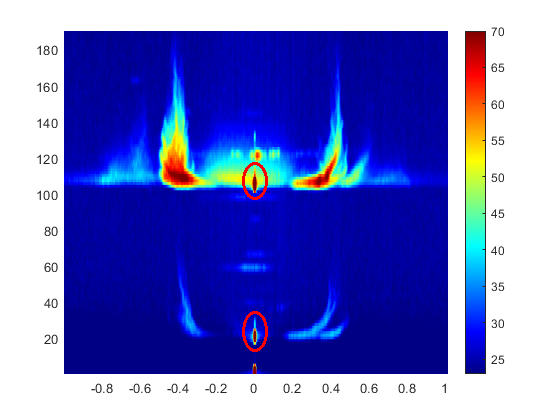}
	\caption{Bistatic Range-Doppler spectra on the Peyras (top panel) and B{\'e}nat (bottom panel) receiver with 2 simultaneous transmitters. The upper floor corresponds to the Porquerolles transmitter and the lower floor to the Peyras transmitter. The direct signal from the transmitter to the receiver is a strong echo at the zero-Doppler cell (red circles)\label{fig04}}

\end{figure}

The HFR network in Toulon is manufactured by WERA Helzel Messtecknik. It is composed of 2 transmitters and 2 receivers located on three distant sites (Figure \ref{fig2}). A standalone transmitter is located on Porquerolles Island, 27 km South-East of Toulon; its single, non-directional, emitting antenna illuminates a wide sea area to the South. A first receiver is located at Cap B\'enat, 35 km East of Toulon, with a regular linear array of 12 receiving active antennas (70 deg from North, anticlockwise) with $0.45$ $\lambda$ spacing. The second transmitter and receiver are located at Fort Peyras about 8 km South West of Toulon. The receiving array is composed of a linear array of 12 passive antennas along the North-South direction with a $0.45 \lambda$ spacing as well. Note that the present combination of 2 TX and 2 RX leads to 3 bistatic pairs and 1 monostatic pair which can be used for the surface current vector reconstruction. We refer to \cite{guerin_radar2019} for a detailed account of the processing of the radar signal in the multistatic mode. Range gating is obtained by the standard frequency-modulated continuous wave (FMCW) HFR technology. The two transmitters POR and PEY send continuous chirp ramps of duration $0.26$ seconds within a frequency band of 100 kHz around the same central frequency $f=16.150$ MHz, allowing for a 1.5 km range resolution. In the standard frequency-modulated continuous wave (FMCW) HF radar technology, the range gating is obtained by binning the received signal in frequency shifts \cite{gurgel2009}. A complex Doppler spectrum is calculated for every single range cell and antenna by Fourier Transform of the recorded voltage time series at the chirp rate. Its squared modulus is referred to as the omnidirectional Power Spectral Density. In order for the receivers to discriminate the signal scattered from the two different sources, the two emitting central frequencies are offset by a small multiple of a frequency bin in such as way that the first half of the range cells are allocated to one transmitter and the other half to the second transmitter. Figure \ref{fig04} shows typical range-Doppler maps obtained by processing the range resolved temporal signal received on a single antenna in Cap B\'enat and Fort Peyras, respectively. The first ``floor'' corresponds to the bistatic or monostatic sea echo from the Fort Peyras transmitter while the second floor is the bistatic return from the Porquerolles transmitter. Note the typical features of the three bistatic Range-Doppler spectra, that is U-shaped Bragg lines and an offset in range corresponding to half the straight distance between transmitter and receiver.

\section{Beam-Forming and its issues}\label{sec:BF}
\begin{figure}[htbp]
\centering
	\includegraphics[scale=0.35]{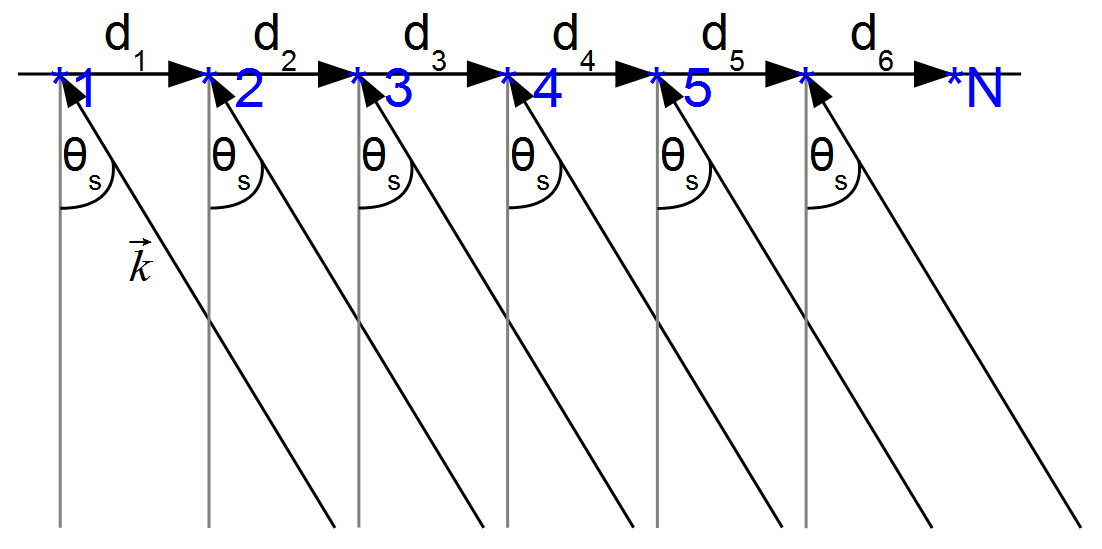}
	\caption{Incident plane wave on a periodic array}
	\label{fig01}
\end{figure}

\begin{figure}[htbp]
\centering
	  \includegraphics[scale=0.4]{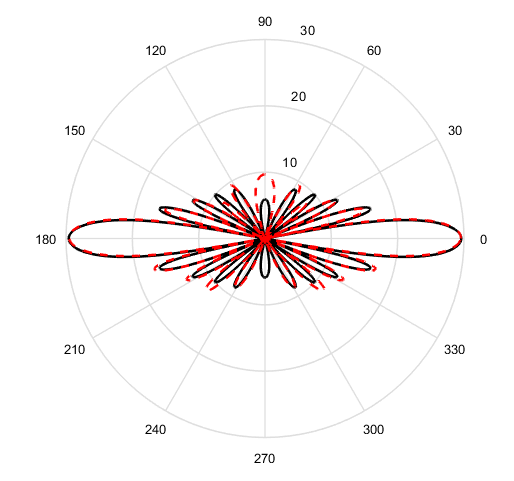}
		\includegraphics[scale=0.4]{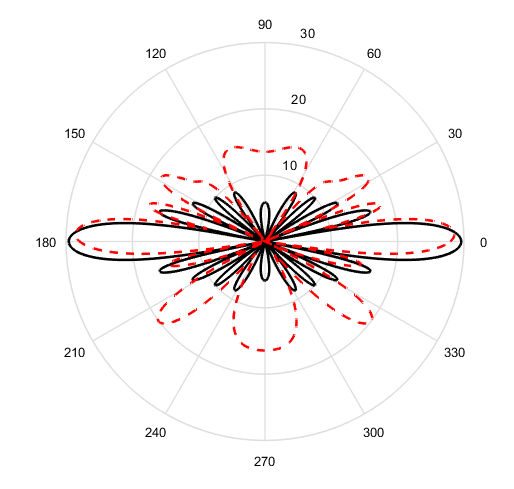}
	\caption{Square modulus of the array factor (in dB) for a 12 antenna linear array with $0.45 \lambda$ spacing with a $\pm 10$ degrees (left panel) and $\pm 50$ degrees (right panel) maximal random perturbation. The theoretical pattern is given in solid black lines.}
	\label{fig02}
\end{figure}

For extended antenna arrays, BF is the traditional method to discriminate the radar signal in azimuth. Consider the canonical problem of a periodic linear array of $N$ identical antennas with spacing $d$, illuminated by a plane wave with wavenumber $K=2\pi/\lambda$ incoming from a direction $\theta_s$ measured from the normal to the array (Figure \ref{fig01}). In the absence of noise and assuming a perfectly coherent monochromatic incident wave, the complex time series recorded on the I\&Q channels of each antenna is proportional to $e^{i 2\pi f t} S_n$, where $f$ is the radar frequency and $S_n$ is a complex antenna gain depending only on the source direction $\theta_s$. For simplicity we assume that all antennas have the same unit gain in amplitude and differ only by a phase shift. Once normalized by the first antenna, the complex signal $S_n$ on the nth antenna depends only on its position in the array and the direction of the source:
\be
S_n=e^{-i(n-1) Kd \sin\theta_s }
\ee
The so-called Array Factor (\cite{balanis2016antenna}):
\begin{equation}
AF(\theta) = \sum_{n=1}^{N} e^{i(n-1)Kd (\sin{\theta}-\sin{\theta_s})}
\end{equation}
gives the localization spot in the far-field that would be produced by an array of identical radiators $S_n$ by virtue of the principle of reciprocity. It is maximal in the direction of the source $\theta=\theta_s$ with a main lobe of width $\lambda/(Nd)$. An array factor can also be defined for irregular arrangements. This is important as the site topography often prevents from installing the complete array along a straight line. Denoting $\vect d_n$ the relative vector position of the nth antenna with respect to the first one, the array factor for an incoming source in direction $\theta_s$ can be written as:
\begin{equation}\label{AF}
AF(\theta) = \sum_{n=1}^{N} e^{+iK (\vect u(\theta)-\vect u(\theta_s))\cdot \vect{d_n}}
\end{equation}
where $\vect u(\theta) $ is the outgoing unit vector in direction $\theta$. For later use note that the array factor can be expressed as a scalar product between the so-called steering vectors in direction of the source ($\theta_s$) and in direction of observation ($\theta$):
\be
AF(\theta) =\vect a(\theta)\cdot\vect a(\theta_s)^\ast,
\ee
where the steering vector is defined as:
\be
\vect a(\theta)= \left(1,e^{iK \vect u(\theta)\cdot \vect{d_1}},...,e^{iK \vect u(\theta)\cdot \vect{d_N}}\right)
\ee
BF consists in re-radiating the received complex multidimensional signal $\vect S=(S_1,...,S_N)$ at infinity while continuously steering the angle of focus $\theta_s$ in the array factor in order to unveil all incoming sources. This is done for each chirp number and range index so that in the end the following complex  time series is obtained as a function of range ($R$), bearing ($\theta$) and time ($t$):
\be\label{beamformingX}
X(R,\theta,t)=\vect a(\theta)\cdot\vect S^\ast(R,t)
\ee
A directional Power Spectral Density can be calculated by squared modulus Fourier Transform of this matrix along the time axis leading to the so-called directional Range-Doppler spectra. The angular sharpness of the AF determines the accuracy and the quality of the azimuthal discrimination and therefore the resolution of the Bragg lines in the directional Doppler spectra. In particular, secondary lobes can be an important source of error if insufficiently rejected since they may ``capture'' strong sources (through their Bragg lines) away from the focusing direction. Secondary lobes can be efficiently rejected by using tapering window but the azimuthal resolution remains bound to the array extension. In addition, the perturbations of the electromagnetic environment as well as the misalignment of the antennas induce phase shifts with respect to the theoretical values which depend only on the array geometry. This results in a deformation of the array factor with possible mispointing and enhancement of secondary lobes. As an example, Figure \ref{fig02} shows the theoretical antenna pattern (solid black lines) for a 12 antenna array along the North-South axis aiming in the Eastward direction. The dashed red plot shows the same pattern after a perturbation of the complex antenna gains by uniform random phases between $-10$ and $+10$ degrees (left panel) or between $-50$ and $+50$ degrees (right panel). In this last case, the phase perturbation induces a mispointing of a few degrees and a strong enhancement of secondary lobes. This will cause a systematic error in the direction of arrival of any source.

\section{Self-calibration}\label{sec:sc}

The classical technique for HFR antenna calibration is based on using a transponder on a boat trip surrounding the emission site and measuring the complex response of each antenna, the so-called antenna manifold. This operation is costly and inaccurate. We instead propose to take advantage of the bistatic configuration to calibrate the antennas using the direct signal of the remote transmitter. The direct signal refers to the EM wave train which propagates in straight line from the transmitter to the receiver without being scattered by sea surface patches on the travel path. It corresponds to the minimal bistatic distance, which is the distance between the transmitting and receiving sites. It is also concentrated on the zero-Doppler cell because the source is not Doppler-shifted by waves. The direct signal is therefore a strong echo concentrated on a particular cell in the range-Doppler map. It can be clearly seen in Figure \ref{fig04} as the red spots circled in red. The complex direct signal $D_n$ recorded on the $n$th antenna can thus be extracted from the zero-Doppler cell at the minimal range after range and Doppler processing of the I and Q signals. Now, from the receiving array point of view, the direct signal is that from an incoming plane wave in direction $\theta_s$ of the transmitter. It should therefore produce the expected relative phase shifts $-K \vect u(\theta_s)\cdot \vect d_n$ ($=-(n-1)Kd\sin\theta_s$ for a linear periodic array). In terms of complex gain this means:

\begin{equation}
D_n=e^{-i K (\vect u(\theta_s) \cdot \boldsymbol{d}_n)}D_1=e^{i\phi_n}D_1
\end{equation}

\begin{figure}[htbp]\centering
  \includegraphics[scale=0.45]{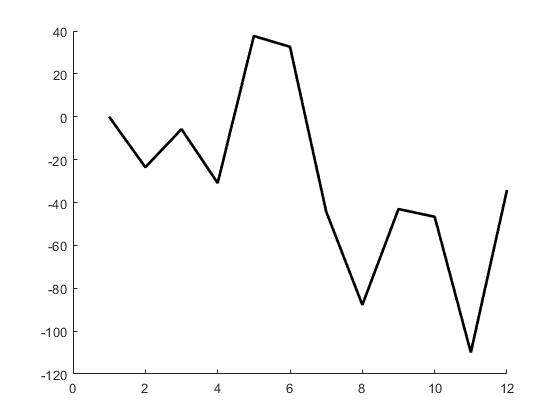}
  \includegraphics[scale=0.45]{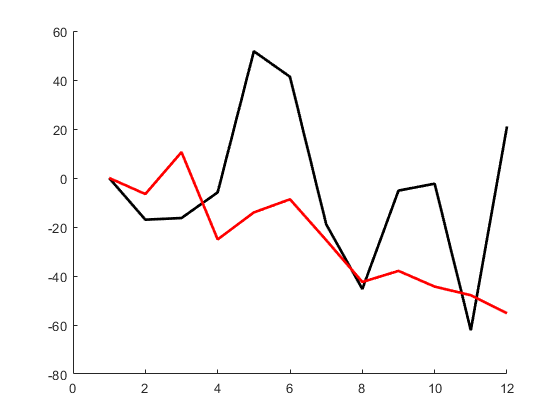}
  \caption{The black solid lines show the required correction of phase, $\delta\phi_n$ (here in degrees), between the measured direct signal and the theoretical phase shifts along the receiving array in Cap B\'enat for a radiation incoming from Porquerolles (left) and Fort Peyras (right) transmitter. The red solid line shows the difference between the measured and corrected theoretical phases in direction of the Fort Peyras transmitter after a self-calibration with the Porquerolles transmitter.\label{fig05}}
\end{figure}

The idea of self-calibration is to compare the theoretical geometrical phase shift between antennas to the actual phase shift measured from the direct signal. The difference between the expected and actual phase shift is a phase perturbation which should be compensated for when processing the antenna signals in azimuth. The self-calibration procedure therefore runs as follows:
\begin{enumerate}
\item For each chirp and each antenna, extract the complex direct signal $\tilde D_n$ on each antenna by retaining the zero-Doppler/minimal range cell from the range-resolved I \& Q signal and calculate its phase $\tilde\phi_n = \arg(\tilde D_n/\tilde D_1)$ relative to the first antenna in the array.
\item Calculate the phase difference $\delta \phi_n= \tilde\phi_n- \phi_n$ with respect to a theoretical array illuminated by a plane wave in direction of the transmitter ($\theta_s$).
\item Before extracting the directional signal $X(R,\theta,t)$ as in (\ref{beamformingX}), correct the complex antenna gain by this phase difference:
  \be
  \tilde S_n(R,t) = S_n(R,t) e^{i \delta \phi_n}
  \ee
\end{enumerate}
By doing this we assume that the phase corrections $\delta \phi_n$ do not depend on the bearing $\theta$. This important property is not granted {\it a priori} and the only certitude is that these phase corrections are appropriate in the direction $\theta_s$ of the transmitter. We therefore need the additional assumption that the required phase corrections $\delta \phi_n$ do not vary (at least, not appreciably) with the bearing. This last hypothesis can be easily tested in direction of the second transmitter where a direct signal can also be extracted and compared with the predicted and corrected phase shifts using the first transmitter. Figure \ref{fig05} (left panel) shows the phase corrections $\delta\phi_n$ obtained along the receiving array of Cap B\'enat in the direction of the two transmitters. For this, the theoretical phase shifts $\phi_n$ have been evaluated from the sole antenna positions and the incoming source direction  while the experimental phase shifts $\tilde\phi_n$ have been extracted from the direct signal with at the zero-Doppler/minimal range cell. In both cases, the phase reference is taken to be that of the first antenna in the array. As seen, the phase difference can be as high as 100 degrees, thereby inducing strong perturbation of the array factor. When correcting the theoretical phase shifts in direction of the second transmitter (Fort Peyras) with the direct signal measured from the first transmitter (Porquerolles), a much better agreement is obtained with the direct signal measured from the second transmitter (Fort Peyras), as seen in the right panel of  Figure \ref{fig05}.

\begin{figure}[htbp]\centering
		\includegraphics[scale=0.5]{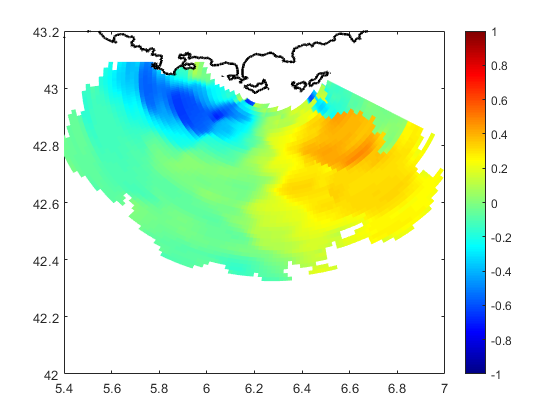}
		\includegraphics[scale=0.5]{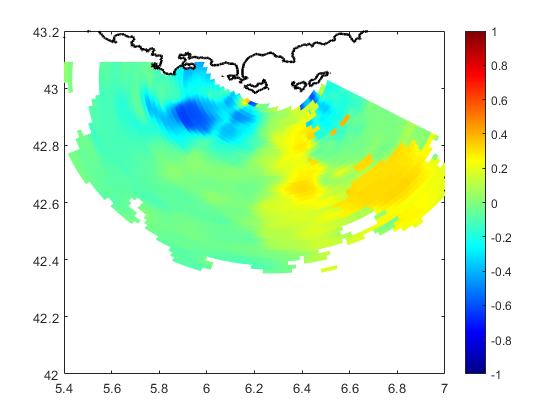}
	        \caption{Elliptical velocity map obtained  on October 13, 2019, 17.00 UTC from one hour observation with the Porquerolles transmitter and Cap B\'enat receiver. The azimuthal processing of HFR data has been performed using the BF method with (left panel) and without (right panel) self-calibration.\label{fig03}}            
\end{figure}

Figure \ref{fig03} shows a map of elliptical velocities obtained  on October 13, 2019, 17.00 UTC, by processing one hour of data recorded in Cap B\'enat and linked to an emission from the Porquerolles transmitter The azimuthal processing has been performed with (left panel) and without (right panel) self-calibration of the complex antenna gains. As plainly seen, this has a dramatic impact on the quality of the resulting map.

\section{Direction Finding and its issues}\label{sec:DF}

\begin{figure}[htbp]\centering
	\includegraphics[scale=0.5]{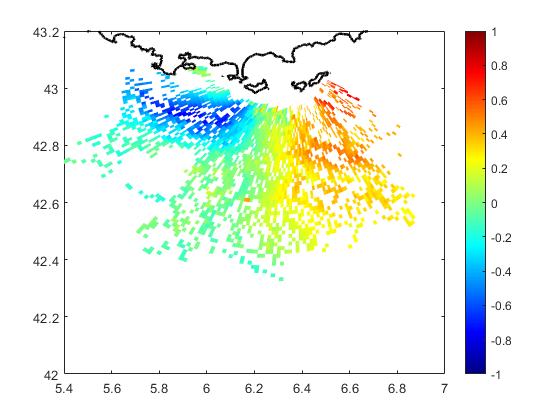}
	\includegraphics[scale=0.5]{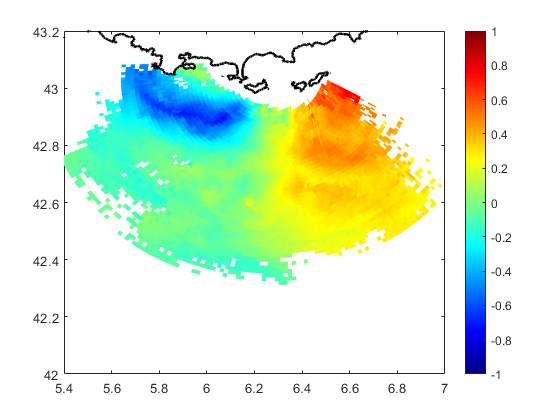}
	\caption{Elliptical velocity obtained  on October 13, 2019, 17.00 UTC from one hour observation with the Porquerolles transmitter and Cap B\'enat receiver. The azimuthal processing of HFR data has been performed using DF azimuthal processing with one single array of 12 antennas (left panel) and antenna grouping with all subarrays of 4 to 12 antennas	(right panel). The self-calibration of antenna phases has been applied in both cases.\label{fig89}}
\end{figure}

The Direction Finding (DF) technique is a high-resolution method for determining the directions of arrival (DOA) of unknown sources in the far field. It is based on the  MUSIC (MUltiple SIgnal Classification) algorithm  \cite{bienvenu1983optimality,schmidtAP86,krim1996two} applied to each complex Doppler ray measured on every single antenna. It allows to find at most $N-1$ DOA by combining $N$ antennas. In the context of surface current mapping the DOA are the bearings associated to each given value of the radial or elliptical velocity, corresponding to a given Doppler shift. The MUSIC algorithm assumes that the signal originating from the different sources is a discrete, stationary random process (of dimension $N$) and is perturbed by an additive white random noise vector. The covariance matrix $\Sigma$ of the complex antenna signals is a $N\times N$ matrix with elements 
\begin{equation}\label{covmatrix}
  \Sigma_{ij} = \mathrm{Cov}\left(Y_i, Y_j^\ast\right)
\end{equation}
where $Y_i $ is the complex Doppler ray (at a given frequency shift) from the ith antenna. To evaluate this quantity, the coherent time series recorded on each antenna channel is split in overlapping intervals and an estimation of the complex Doppler spectrum is obtained for each time interval. This provides for each pair $(i,j)$ of antennas and each Doppler ray a certain number of (quite) independent samples from which the ensemble average in (\ref{covmatrix}) can be evaluated. The optimal number of samples results from a trade-off between the convergence of the matrix cross-products to their statistical mean and the Doppler frequency resolution which decreases with the sample size.
Next a Singular Value Decomposition of the covariance is sought, $\Sigma = U \Lambda U^\ast$ where $\Lambda=\mathrm{Diag}(\lambda_1,..., \lambda_N)$ is the diagonal matrix of singular values ($\lambda_1\geq ...\geq\lambda_N\geq 0$) and $U=\left(\boldsymbol{U}_1,...,\boldsymbol{U}_N\right)$ is the matrix of eigenvectors $\boldsymbol{U}_j$. In the absence of noise, the $M$ positive singular values $\lambda_1\geq ..\geq\lambda_M>0$ identify the number and  strength of the different sources and the associated eigenvectors  $\boldsymbol{U}_1,..,\boldsymbol{U}_M$ their direction. Precisely, the eigenvector coincides with the normalized steering vector in the direction $\theta_j$ of the source, $\boldsymbol{U}_j=\boldsymbol{a}(\theta_j)/\norme{\boldsymbol{a}(\theta_j)}$. The remaining $N-M$ eigenvalues $\lambda_{M+1}=..=\lambda_N= 0$ and associated  eigenvectors  $\boldsymbol{U}_{M+1}..\boldsymbol{U}_N$ define the null subspace. In the presence of noise, these last eigenvalues are actually nonzero but are supposed to be clearly lower than the signal eigenvalues ($\lambda_{N}\leq ..\leq\lambda_{M+1}<<\lambda_{M}$) and the corresponding subspace is called the noise subspace. The idea of the MUSIC algorithm is to identify the direction of arrival by minimizing the projection of the steering vector onto the noise subspace, which amounts to maximize its inverse:

\begin{equation}
Q(\theta) = \frac{|| \boldsymbol{a}(\theta) ||}{||\sum_{m=M+1}^{N} [ \boldsymbol{a}(\theta) \cdot \boldsymbol{U^*}_m ] \boldsymbol{U}_m  ||}
\end{equation}
This quantity is referred to as the ``MUSIC factor''; its first $M$ maxima above some threshold identify the sources. The main advantage of the DF method with respect to the BF method is an improved azimuthal resolution. The actual obtained resolution with DF is not clearly defined but is somehow related to the width of the MUSIC factor and the size of the antenna array. However, the main shortcoming of DF when employed with a large number of antennas is to produce lacunary maps. This can be due to either an insufficient number of sources of an insufficient number of bearings reaching the MUSIC threshold. The left panel in Figure \ref{fig89} shows an example of elliptical velocities obtained with the Porquerolles-Cap B\'enat pair on the same date as previously (October 13, 2019, 17:00 UTC). The range-resolved time series have been processed with DF using the full array (12 antennas) and assuming 3 sources. As seen, the azimuthal resolution is greatly improved with respect to the BF technique. However, the map is very lacunary as many bearing were not identified. The filling ratio of the map depends very much on the chosen threshold for the MUSIC factor. The choice of the latter results from a trade-off between lacunarity and erroneous detection of fictitious sources.

\section{Antenna grouping}\label{sec:ag}

A major improvement to prevent the lacunarity of radial or elliptical surface current maps is to extend the DF processing to all possible combinations of subarrays made of consecutive antennas, a method which we refer to as ``antenna grouping''. Instead of restricting the covariance analysis to a single maximal set of antennas, we apply it as many times as there are admissible subarrays of arbitrary size. The idea of using subarrays in DF is not completely new as it is a well-known method (see the review in \cite{krim1996two}) to improve the estimation of the covariance matrix at the cost of a reduction of the signal subspace dimension. However, we take advantage of subarrays in a different way, as we use them to obtain many independent coarse estimations of elliptical velocities and allow them an arbitrary size, from a minimal $N_{min}$ (typically, 3 or 4) to a maximal size $N$ corresponding to the length of the full array. It is simple combinatorics to see that there are $(N-N_{min}+1)(N-N_{min}+2)/2$ such subarrays. The idea is to perform a weighted mean of elliptical velocities obtained with the different subsets of antennas, by averaging all the elliptical velocities falling into each bin of bearing. As they are many subarrays, this augments considerably the probability of visiting a given radar cells (that is, a range and bearing) with the DF algorithm. In fact most bearings will be visited several times while looping over the subarrays, with possibly different values of the elliptical velocity (that is, possibly distinct Doppler rays). This improves drastically the filling of the map and also provide more reliable and accurate estimates of elliptical velocities with less outliers. Smaller groups of antenna have a limited azimuthal resolution. As a result, they are less accurate to evaluate the surface elliptical velocity because they tend to smooth the latter. In addition, they are limited to a small number of sources with the effect that they can miss some specific features of the surface current pattern in case of complex meandering structures. However they are more robust to noise and have a better filling factor. On the other hand, larger groups have an increased azimuthal resolution and accuracy and allow for a larger number of sources but are more lacunary. These complementary strengths and weaknesses of small and large antenna groups mitigate each other when the outcome of all possible subarrays are averaged. A stronger weight is attributed to the current estimated with large antenna groups which are sparser but more accurate. The chosen weights as well as the chosen MUSIC thresholds and the number of sources for each application of the DF on subarrays will not be given here as this depends on each particular radar system and must be optimized beforehand. The right panel in Figure \ref{fig89}  shows the same map as previously with a weighted mean of the elliptical velocities obtained with a DF process applied to consecutive groups of $4$ to $12$ antennas (45 combinations) with a preliminary complex antenna gain correction with the self-calibration method. A remarkable increase of the spatial coverage is obtained with almost the same filling rate as with BF while it is obvious that the azimuthal resolution has been drastically improved with respect to the latter.

\begin{figure}[htbp]\centering
	\includegraphics[scale=0.5]{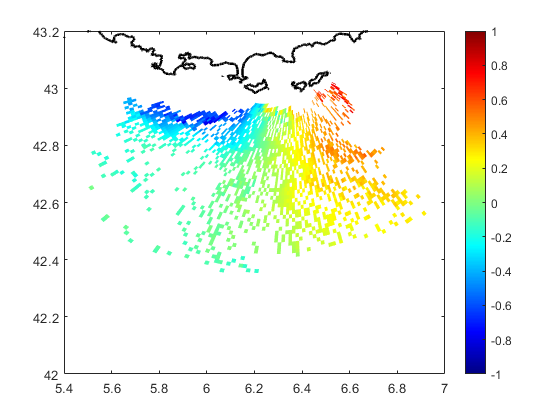}
	\includegraphics[scale=0.5]{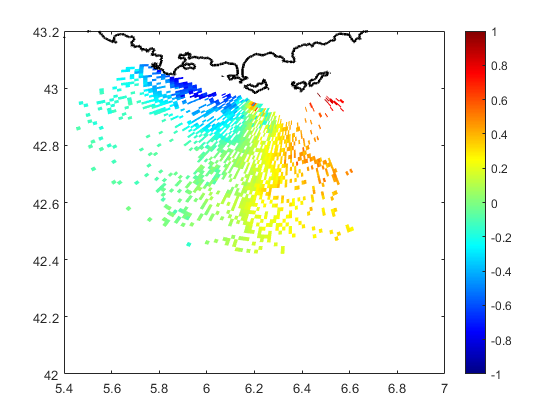}
	\caption{Same case as Figure \ref{fig89} with DF azimuthal processing using one single subarrays of 3 antennas in the absence of self-calibration. Two different subarrays have been used to generate the left and right panels. A rotation is visible between the two maps as an artifact of the antenna miscalibration.\label{fig10}}
\end{figure}

\begin{figure}[htbp]\centering
	\includegraphics[scale=0.5]{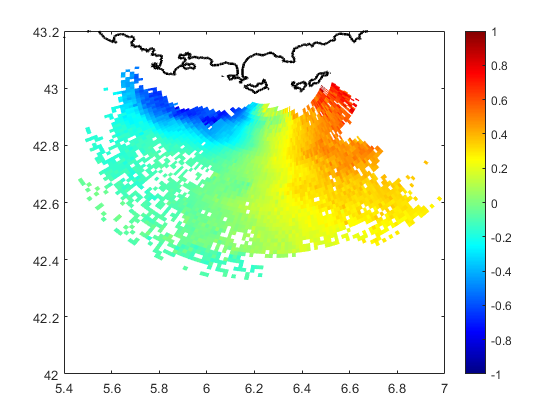}
	\includegraphics[scale=0.5]{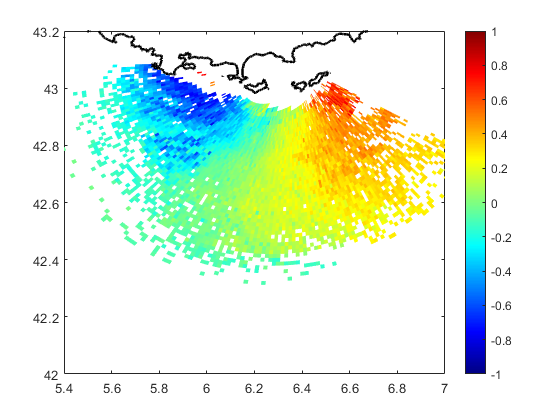}
	\caption{Same case as Figure \ref{fig89} with DF azimuthal processing using 10 admissible subarrays of 3 antennas with (left panel) and without (right panel) self-calibration.\label{fig11}}

\end{figure}

The key point in using small antenna subarrays is a good preliminary calibration which is a requisite for the success of antenna grouping. Whereas a missing or imperfect calibration of antenna gains is less visible when performing DF on a long array, it can have a drastic impact on short arrays. As seen in Figure \ref{fig10}, a miscalibration of the complex antenna gains produces a small rotation of the elliptical current map. When averaging the individual elliptical currents obtained with the different short arrays, this results in a blurring of the final elliptical velocity map. This artifact can be mitigated with the self-calibration technique which allows to correct the complex antenna gains first inferred from geometrical considerations. To test the efficiency of self-calibration in this regards, we produced surface current maps with 10 groups of 3 antenna with and without self-calibration (Figure \ref{fig11}). The gain in quality is quite obvious.

\begin{figure}[htbp]\centering
  \includegraphics[scale=0.5]{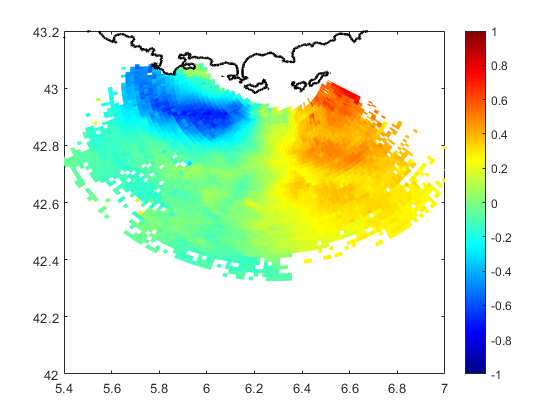}
	\includegraphics[scale=0.5]{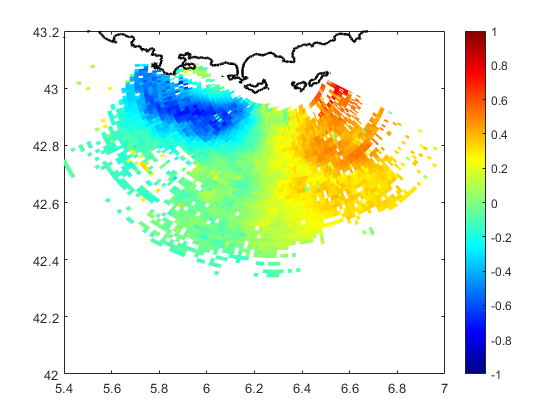}
	\caption{Same case as Figure \ref{fig89} with DF azimuthal processing using antenna grouping with all subarrays of 4 to 12 antennas whenever 1 antenna (number 7, left panel) or 3 antennas (number 3,7 and 8, right panel) in the array are out of service.\label{fig12}}
\end{figure}

Besides increased resolution and filling ratio, another striking performance of the antenna grouping method is its robustness to the failure of some antennas in the array. This is a very valuable quality as it often occurs for various reasons that one or several antennas in the remote receiving sites are out of service (because of e.g. corrosion, damaged individual receiver or connector, vandalism, etc) and cannot be repaired at once. Contrarily to BF, the DF with antenna grouping can still be efficient with an incomplete array, even though the resulting elliptical current maps deteriorate little by little as an increasing number of antennas are missing. Figure \ref{fig12} shows the same radial current map as in Figure \ref{fig89} obtained by applying the DF method with antenna grouping to all subarrays of 4 to 12 antenna whenever a certain number of antennas are discarded. 

\section{Conclusion}
The bistatic configuration offers the possibility to calibrate the antenna arrays automatically by adjusting the geometrical phases in direction of the transmitter to the actual phase shift measured with the direct signal which can be extracted from the received data with the zero-Doppler method. This correction of the complex antenna gains yields significant improvement in the azimuthal processing both with BF and DF methods. We also found that DF can be successfully employed for linear arrays when applied simultaneously to all existing subarrays, a technique which we referred to as antenna grouping. When combined with the self-calibration method it leads to high-resolution maps of elliptical velocity with full coverage. We presented here the main principles of the improved mapping and did not provide the numerical values of the group weights and MUSIC parameters (MUSIC factor thresholds and number of sources) that have been chosen in the azimuthal processing of our data. There is no universal recipe to set these parameters which must be tuned and optimized for each radar system and are part of the ``know-how'' of each operator. The validity and accuracy of our procedure for surface current extraction has been validated recently with several launches of drifters in the area covered by the radar (\cite{dumas_OD20}). For one type of drifter that integrates the surface current over the 65 cm top layer, we found an excellent agreement between the in situ measurements and the HFR derived elliptical velocities, of the order of 2.5 cm/s on the best trajectories. This will be confirmed by further oceanographic campaigns.

{\small\paragraph{Acknowledgments}
  The upgrade and maintenance of the WERA HF radar system in Toulon have been funded by the EU Interreg program SICOMAR-PLUS and trusted to the company Degreane Horizon over the period 2019-2021. We acknowledge the University of Toulon and the Minist\`ere de l'Enseignement Sup\'erieur, de la Recherche et de l'Innovation for funding the first author. We thank the Parc National de Port-Cros (PNPC) for its support and hosting of our radar transmitter in Porquerolles Island. We also thank the ``Association Syndicale des Propri\'etaires du Cap B\'enat'' (ASPCB) for allowing our receiver array at the Cap B\'enat as well as the Group Military Conservation and the Marine Nationale for hosting our radar installation in Fort Peyras. }


\begin{thebibliography}{10}

\bibitem{crombie_Nature55}
D~D Crombie.
\newblock Doppler spectrum of sea echo at 13.56 mc./s.
\newblock {\em Nature}, 175(4459):681--682, 1955.

\bibitem{barrick_72conf1}
D~E Barrick.
\newblock Remote sensing of sea state by radar.
\newblock In {\em Engineering in the Ocean Environment, Ocean 72-IEEE
  International Conference on}, pages 186--192. IEEE, 1972.

\bibitem{lipa1981}
BJ~Lipa, DE~Barrick, and JW~Maresca~Jr.
\newblock Hf radar measurements of long ocean waves.
\newblock {\em Journal of Geophysical Research: Oceans}, 86(C5):4089--4102,
  1981.

\bibitem{paduan1997}
J.D. Paduan and H.C. Graber.
\newblock Introduction to high-frequency radar: reality and myth.
\newblock {\em Oceanography}, 10(2):36--39, 1997.

\bibitem{headrick1998}
JM~Headrick and JF~Thomason.
\newblock Applications of high-frequency radar.
\newblock {\em Radio Science}, 33(4):1045--1054, 1998.

\bibitem{paduan2013}
J.D. Paduan and L.~Washburn.
\newblock High-frequency radar observations of ocean surface currents.
\newblock {\em Annual review of marine science}, 5:115--136, 2013.

\bibitem{wyatt2014}
LR~Wyatt.
\newblock High frequency radar applications in coastal monitoring, planning and
  engineering.
\newblock {\em Australian Journal of Civil Engineering}, 12(1):1--15, 2014.

\bibitem{roarty2019}
H.~Roarty et~al.
\newblock The global high frequency radar network.
\newblock {\em Frontiers in Marine Science}, 6:164, 2019.

\bibitem{grosdidier_GRS14}
S.~Grosdidier, P.~Forget, Y.~Barbin, and C.-A. Gu{\'e}rin.
\newblock {HF} bistatic ocean {D}oppler spectra: Simulation versus
  experimentation.
\newblock {\em IEEE Trans. Geosci. and Remote Sens.}, 52(4):2138--2148, 2014.

\bibitem{dumas_OD20}
D.~Dumas, A.~Gramoull{\'e}, C.-A. Gu{\'e}rin, A.~Molcard, and Y.~Ourmi{\`e}res
  \and B.~Zakardjian.
\newblock Multistatic estimation of high-frequency radar surface currents in
  the region of {T}oulon.
\newblock {\em Ocean Dynamics}, 2020.
\newblock submitted.

\bibitem{guerin_radar2019}
C.-A. {Gu{\'e}rin}, D.~{Dumas}, A.~{Gramoull{\'e}}, C.~{Quentin},
  M.~{Saillard}, and A.~{Molcard}.
\newblock The multistatic oceanographic {HF} radar network in {T}oulon.
\newblock In {\em 2019 International Radar Conference (RADAR)}, pages 1--5,
  2019.

\bibitem{barbin09}
Y.~Barbin, P.~Broche, and P.~Forget.
\newblock {High Resolution Azimutal Radial Current Mapping with Multisource
  Capability}.
\newblock In {\em {Radio Oceanography Workshop 2009}}, Split, Croatia, May
  2009.

\bibitem{barbin11}
Y.~Barbin.
\newblock {High Resolution Surface Currents Mapping using Direction Finding
  Method in Bistatic Radar Configuration}.
\newblock In {\em {Third Conference on Remote Ocean Sensing}}, La Spezia,
  Italy, October 2011.

\bibitem{gurgel2009}
K.-W. Gurgel and T.~Schlick.
\newblock Remarks on signal processing in {HF} radars using {FMCW} modulation.
\newblock In {\em Proc. IRS}, pages 1--5, 2009.

\bibitem{balanis2016antenna}
Constantine~A Balanis.
\newblock {\em Antenna theory: analysis and design}.
\newblock John wiley \& sons, 2016.

\bibitem{bienvenu1983optimality}
G.~Bienvenu and L.~Kopp.
\newblock Optimality of high resolution array processing using the eigensystem
  approach.
\newblock {\em IEEE Transactions on acoustics, speech, and signal processing},
  31(5):1235--1248, 1983.

\bibitem{schmidtAP86}
Ralph Schmidt.
\newblock Multiple emitter location and signal parameter estimation.
\newblock {\em IEEE transactions on antennas and propagation}, 34(3):276--280,
  1986.

\bibitem{krim1996two}
H.~Krim and M.~Viberg.
\newblock Two decades of array signal processing research.
\newblock {\em IEEE signal processing magazine}, 1996.

\end{thebibliography}

\end{document}